\documentclass[preprint,12pt]{elsarticle}

\usepackage{amssymb}
\usepackage{amsmath}
\usepackage{graphicx}
\usepackage{subcaption}
\usepackage{xcolor}
\usepackage[normalem]{ulem}

\journal{Journal of Subatomic Particles and Cosmology}

\begin{document}

\begin{frontmatter}



\title{Investigating Universal Relations in Compact Stars featuring $\Delta-$Admixed Exotic Dense Matter}


\author{Vivek Baruah Thapa$^{a}$}
\ead{vivek.thapa@bacollege.ac.in}
\affiliation{organization={Department of Physics, Bhawanipur Anchalik College},
            city={Barpeta},
            postcode={781352}, 
            state={Assam},
            country={India}}

\author{Anil Kumar$^b$}
\affiliation{organization={Indian Institution of Technology Jodhpur},
            city={Jodhpur},
            postcode={342030}, 
            state={Rajasthan},
            country={India}}

\author{Vishal Parmar$^c$}
\affiliation{organization={INFN},
            city={Sezione di Pisa, Largo B. Pontecorvo 3},
            postcode={I-56127},
            state={Pisa},
            country={Italy}}

\author{Monika Sinha$^b$}


\begin{abstract}
The dense material in a compact star from a supernova remnant is beyond terrestrial experimentation, so phenomenological modeling is used to match astrophysical observations. This is crucial due to the complex sensitivity of compact star features to dense matter properties. Despite modeling flexibility, certain universal relationships among compact star features hold true, regardless of the matter model. Our study examines these universal relationships, focusing on the moment of inertia, tidal Love number, and quadrupole moment, as well as correlations between non-radial oscillation frequencies and star compactness. We consider baryonic stars with cores of heavier baryons. Our findings show that baryonic stars with cores of heavier baryons follow the universal relations, and the f-mode oscillation frequency's universality relative to tidal deformability is notable, with an error margin under 1$\%$.
\end{abstract}




\end{frontmatter}



\section{Introduction}
\label{sec:intro}

Massive stars end their evolutionary journey in spectacular core-collapse supernovae, leading to the formation of highly compact stellar remnants, commonly identified as neutron stars (NS) or, in some cases, hypothetical quark stars. These objects exhibit an average density of approximately $10^{14}$ gm/cm$^3$, vastly surpassing the nuclear saturation density ($n_0$), beyond which conventional nuclear matter transitions into an ultradense state governed by complex quantum chromodynamic (QCD) interactions. Such extreme conditions render compact stars invaluable as astrophysical laboratories, providing a unique setting to probe the fundamental properties of dense nuclear matter—properties that remain inaccessible to terrestrial experiments due to technological and energy constraints.

Despite extensive theoretical and observational efforts, the precise composition of matter at supranuclear densities remains a fundamental open question in nuclear and astrophysical sciences. The interactions between particles in these environments are highly uncertain, making the equation of state (EOS)-which defines the relationship between energy density and pressure—one of the least constrained aspects of NS physics. A wide array of phenomenological models has been proposed to describe the possible compositions and phase transitions occurring within these stars. The most conventional description assumes purely nucleonic matter, primarily composed of neutrons, with a small fraction of protons and a corresponding population of leptons (electrons and muons) to maintain charge neutrality and beta equilibrium. However, at densities exceeding a few times 
$n_0$, it is theoretically expected that more exotic degrees of freedom may emerge, fundamentally altering the microphysics of the stellar interior.

A significant body of research has explored the viability of various exotic states of matter in NSs. One possibility is the appearance of hyperons—strange baryons such as $\Lambda$, $\Sigma$, and $\Xi-$ which become energetically favorable at high densities, thereby modifying the EOS and softening it \cite{1991PhRvL..67.2414G, 2013PhRvC..87e5806C, 2015JPhG...42g5202O, 2018MNRAS.475.4347R, 2018EPJA...54..133L, 2021NuPhA100922171L}. Similarly, investigations have considered the presence of non-strange baryonic resonances, such as the $\Delta-$resonance states, which can significantly impact the maximum mass and tidal deformability of NSs \cite{2014PhRvC..90f5809D, 2015PhRvC..92a5802C, 2018PhLB..783..234L, 2019ApJ...874L..22L, 2023ApJ...944..206L, 2020arXiv201000981B, 2021MNRAS.507.2991T, PhysRevD.103.063004}. Beyond baryonic matter, mesonic condensations, such as kaon or pion condensates, have been proposed as viable alternatives capable of softening the EOS at intermediate densities while leading to distinct observational signatures, including modifications in neutrino emissivity and gravitational wave signals \cite{2019Parti...2..411M, 1982ApJ...258..306H, 1999PhRvC..60b5803G, 2001PhRvC..63c5802B, 2020PhRvD.102l3007T, 2021arXiv210208787B, PhysRevD.103.063004}.
At even more extreme densities, where the Fermi momenta of baryons become comparable to the QCD scale, a transition to deconfined quark matter is anticipated. This hypothesis is supported by high-energy QCD calculations and models such as the MIT bag model \cite{PhysRevD.9.3471,1984PhRvD..30.2379F} and the Nambu–Jona-Lasinio (NJL) \cite{2005PhR...407..205B,PhysRevLett.37.8,Nambu:1961tp} framework. The presence of a quark-hadron transition within NSs could explain observed anomalies in NS mass-radius relationships, as well as the existence of massive stars ($> 2~M_\odot$), which challenge purely nucleonic models. Additionally, recent advances in multi-messenger astronomy, particularly gravitational wave observations from binary neutron star (BNS) mergers (e.g., GW170817) \cite{PhysRevLett.119.161101}, have begun placing constraints on the tidal deformability of NSs, offering indirect insights into the possible existence of exotic matter in their interiors.
Future observations from next-generation telescopes, such as the Square Kilometer Array (SKA) \cite{2009pra..confE..58L} and the Neutron Star Interior Composition Explorer (NICER) \cite{2016SPIE.9905E..1HG}, combined with advancements in heavy-ion collision experiments at facilities like FAIR (GSI) \cite{SPILLER2006305} and NICA (JINR) \cite{Kekelidze_2017}, are expected to provide crucial data for refining our understanding of dense matter. Ultimately, resolving the composition of matter inside NSs is not only fundamental to nuclear astrophysics but also essential for understanding the nature of strong interactions under extreme conditions.

Extracting all global properties of a NS simultaneously from observations is generally infeasible due to observational limitations and inherent degeneracies in astrophysical measurements. Consequently, a crucial aspect of NS research is the identification of empirical relationships that connect only a few key physical quantities. These relationships are typically dependent on the EOS and include fundamental correlations such as the mass-radius relation of NSs. By accurately measuring these quantities, astrophysicists aim to place meaningful constraints on the EOS, thereby improving our understanding of ultra-dense matter. However, from a fundamental physics perspective, it is even more compelling to investigate whether certain relationships exhibit universality—remaining largely independent of the underlying EOS models. Such EOS-insensitive relationships could serve as powerful tools for probing gravitational theories, allowing for stringent tests of general relativity in strong-field regimes despite the persistent uncertainties surrounding the microphysics of dense nuclear matter \cite{2013PhRvD..88b3009Y, 2013Sci...341..365Y}.  
A significant breakthrough in this domain was the discovery of universal relations among key macroscopic properties of NSs. Specifically, Refs. \cite{2013PhRvD..88b3009Y, 2013Sci...341..365Y} identified a set of EOS-independent relations linking the moment of inertia (\(I\)), the quadrupole moment (\(Q\)), the tidal Love number (\(\lambda\)), and the rotational Love number (\(\lambda_{\rm{rot}}\)). Collectively referred to as the \(I\)-Love-\(Q\) relations, these correlations have profound implications for gravitational wave astronomy, as they enable the extraction of NS properties from gravitational wave signals detected during the late inspiral and merger phases of BNS coalescences. Since these relations do not strongly depend on the underlying EOS, they provide a robust method for testing deviations from general relativity in extreme astrophysical environments.  

Over the past decade, numerous studies have introduced additional empirical formulae that establish relationships between different physical characteristics of NSs. For instance, early works \cite{2002A&A...396..917B, 2005ApJ...629..979L} uncovered correlations between the ratio of \(I\) to the mass and squared radius (\(I/MR^2\)) and the compactness parameter (\(M/R\)). These relations offer an alternative approach to indirectly infer NS structure and constrain its EOS from limited observational data. Furthermore, Ref. \cite{2013MNRAS.433.1903U} demonstrated the existence of a universal relationship between the ratio of the spin-induced \(Q\) and the square of the angular momentum (\(J^2\)) with the compactness, further strengthening the case for EOS-independent descriptions of NS properties.  
Beyond purely nucleonic models, a growing body of research has explored the impact of exotic matter on universal relations. The presence of strange baryons, such as hyperons, has been investigated in Ref. \cite{2018EPJA...54...26B}, where their inclusion alters NS structure while still preserving certain empirical correlations. Similarly, studies on non-strange baryonic resonances, including \(\Delta\)-baryons, have demonstrated their effects on NS deformation and stability \cite{2020MNRAS.499..914R}. Additionally, deconfined quark matter—expected to emerge at extreme densities—has been considered in several works \cite{2022MNRAS.515.3539K, 2018PhRvD..97h4038P}, revealing that quark degrees of freedom can introduce deviations in universal relations, though some level of EOS insensitivity remains.  

These investigations collectively highlight the profound significance of universal relations in NS astrophysics. Not only do they offer a means to infer NS properties from limited observational data, but they also provide a crucial framework for distinguishing between competing EOS models and testing the validity of general relativity in the strong-field regime. With the advent of next-generation gravitational wave detectors, X-ray timing missions such as NICER, and advancements in multi-messenger astronomy, these universal relations are expected to play a pivotal role in unraveling the fundamental nature of dense matter and the laws of gravity governing the most extreme astrophysical objects.

Asteroseismology provides yet another crucial perspective on universal relations in NSs, offering insights into their internal structure and composition through the study of non-radial oscillation modes. These oscillation modes are classified based on their respective restoring forces: fundamental (\(f\)-) modes, pressure (\(p\)-) modes, gravity (\(g\)-) modes, and space-time (\(w\)-) modes. The study of these modes is particularly important as they serve as potential sources of gravitational waves, allowing for the indirect probing of NS interiors via multi-messenger observations.  
Recent studies \cite{2022PhRvD.106f3005K, zhaoUniversalRelationsNeutron2022, lozanoTemperatureEffectsCore2022} have explored the influence of different EOSs on these non-radial oscillation modes, aiming to establish empirical relationships that could help constrain NS properties. In particular, Ref. \cite{thapaFrequenciesOscillationModes2023} investigated the impact of nuclear saturation parameters and finite-temperature EOSs on oscillation frequencies, highlighting how variations in microphysics influence mode characteristics. Such analyses are vital for understanding how NSs evolve under extreme conditions, including during binary mergers where oscillations play a significant role in post-merger gravitational wave signatures.  

A key empirical relation for the \(f\)-mode oscillation frequency (\(\omega_f\)) was first proposed in Ref. \cite{1998MNRAS.299.1059A} using Newtonian theory. It was demonstrated that \(\omega_f\) exhibits an almost linear correlation with the square root of the average density in full general relativity, implying a direct link between the fundamental mode frequency and the global properties of NSs. Additionally, a formula for the damping time (\(\tau\)) due to gravitational wave emission was derived, providing an essential tool for estimating the stability of oscillatory modes in compact stars. Extending these findings, Ref. \cite{2004PhRvD..70l4015B} incorporated updated EOSs, refining the initial empirical fits and improving the accuracy of the predictions.  
Building upon this work, Ref. \cite{2010ApJ...714.1234L} introduced a semi-universal relation involving the dimensionless oscillation frequency, defined as \(\Bar{\omega}_f = GM\omega_f/c^3\), and the dimensionless moment of inertia, given by \(\tilde{I} = Ic^4/(G^2M^3)\). This relation provides a more general framework for describing NS oscillations, offering potential EOS-independent constraints that could be tested with future gravitational wave observations.  
The study of NS oscillations and their associated universal relations is particularly timely in light of recent gravitational wave detections, such as GW170817, which have opened new avenues for probing NS interiors through their dynamic behavior. With next-generation observatories like the Einstein Telescope and Cosmic Explorer on the horizon, the empirical relations governing oscillation modes are expected to become even more critical in deciphering the physics of dense nuclear matter and testing alternative theories of gravity.

The primary objective of this study is to investigate universal relations for non-radial oscillations in NSs while incorporating exotic components in the particle spectrum. Given the possibility of exotic degrees of freedom—such as hyperons, meson condensates, or deconfined quark matter—at extreme densities, understanding how these components influence oscillation modes is crucial for refining astrophysical constraints on the EOS.  
This paper is structured as follows: Sec. \ref{sec:formalism} presents a detailed description of the EOS models considered in this study, along with the theoretical framework employed for their construction. Additionally, a concise overview of non-radial oscillation modes is provided in the same section. The main results, including the established universal relations, are discussed in Sec. \ref{sec:results}, followed by the key conclusions in Sec. \ref{sec:summary}.  
Throughout this work, we adopt natural units (\(\hbar = c = 1\)) for consistency in the relativistic formulations. The study employs both the relativistic Cowling approximation and the full general relativistic approach to compute the quasinormal oscillation modes. However, the results obtained using the Cowling approximation are not included in this paper, as the focus remains on the fully relativistic treatment of oscillations to ensure higher accuracy in the derived universal relations.

\section{Formalism} \label{sec:formalism}
The subsequent section provides a comprehensive discussion on the dense matter EOS models and the Covariant Density Functional framework utilized for their construction. Additionally, it outlines the methodology employed to analyze non-radial oscillation modes, ensuring a systematic approach to evaluating their dependence on different EOS models.

\subsection{Covariant Density Functional Model}
This section provides a detailed description of the Covariant Density Functional (CDF) model framework employed for constructing the dense matter EOS. The CDF approach is widely used in nuclear astrophysics to describe the properties of dense nuclear matter in a relativistically consistent manner, ensuring that key nuclear many-body interactions are systematically incorporated.  

In this study, the composition of matter includes the full baryon octet, consisting of nucleons (\(N\))—protons (\(p\)) and neutrons (\(n\))—as well as hyperons (\(Y\)), which include \(\Lambda\), \(\Sigma\), and \(\Xi\) baryons. Additionally, we extend the particle spectrum by considering the presence of heavier non-strange \(\Delta\)-resonances (\(\Delta^{++, +, 0, -}\)), which can play a crucial role in modifying the EOS at supranuclear densities. The inclusion of these additional baryonic degrees of freedom is essential for a realistic modeling of dense matter, as it can significantly influence the stiffness of the EOS and consequently affect the mass-radius relation of NSs.  
To maintain \(\beta\)-equilibrium and charge neutrality, we also incorporate leptons, specifically electrons (\(e^-\)) and muons (\(\mu^-\)), which act as charge carriers in response to the weak interaction processes occurring within the star. Their presence ensures that the chemical potentials of the different particle species satisfy equilibrium conditions, thereby governing the overall composition of NS matter.  

The inter-particle interactions within the baryonic medium are mediated by mesonic exchange fields, which effectively model the strong force. In our approach, we consider the exchange of the fundamental mesons \(\sigma\) meson (isoscalar-scalar), \(\omega\) meson (isoscalar-vector) and \(\rho\) meson (isovector-vector).
Additionally, to account for interactions involving strange baryons, we introduce the \(\phi\) meson (isoscalar-vector), which mediates the repulsive interaction between hyperons. The inclusion of this meson plays a critical role in determining the onset density and abundance of hyperons in NS cores, thereby influencing the overall EOS and maximum mass of NSs.  

The full theoretical formulation, including the Lagrangian density describing nucleonic, hyperonic, and \(\Delta\)-admixed hypernuclear matter, as well as the derivation of the corresponding EOS within the CDF framework utilized in this study, can be found in Ref. \cite{2021MNRAS.507.2991T}. This formalism provides the necessary foundation for investigating the structural and oscillatory properties of NSs, particularly in the presence of exotic degrees of freedom.

In this study, we employ a set of density-dependent coupling parameterizations, specifically DD-ME2 \cite{2005PhRvC..71b4312L}, DD-MEX \cite{TANINAH2020135065}, and DD2 \cite{PhysRevC.81.015803}, within the framework of the CDF theory. These parameterizations are systematically calibrated to reproduce empirical nuclear matter properties at the saturation density, ensuring a reliable description of dense matter in NSs.  
The DD-ME2, DD-MEX, and DD2 parameter sets have been extensively used in nuclear astrophysics due to their capability to provide a unified description of both finite nuclei and infinite nuclear matter. They incorporate density-dependent meson-baryon couplings, which allow for a more accurate modeling of the nuclear interaction over a wide range of densities encountered in NS interiors. The choice of these parameterizations is motivated by their ability to reproduce key nuclear matter saturation properties, such as the binding energy per nucleon ($E_0$), incompressibility ($K_0$), Dirac nucleonic effective mass ($m^*$), symmetry energy ($E_{\rm{sym}}$), and its slope ($L_{\rm{sym}}$) (please refer to Table - \ref{tab:2}), which are crucial for determining the dense matter EOS and subsequent NS structure solutions.
It is noteworthy to mention that the coupling parameter sets used in this work satisfy the constraints from finite nuclei data as well as model data from various heavy-ion collisions \cite{2017RvMP...89a5007O}.
For a comprehensive understanding of the coupling constants and their specific values used in this study, readers may refer to Ref. \cite{PhysRevC.107.035807}, where these same coupling models were employed to explore the properties of antikaon-condensed matter under extreme magnetic fields. This work serves as a useful reference for understanding the impact of different EOS models on the behavior of dense matter under astrophysical conditions.  
Additionally, for the meson-baryon couplings involving the exotic particles—such as hyperons and \(\Delta\)-resonances—considered in our analysis, we adopt the parameter choices from Ref. \cite{2021MNRAS.507.2991T}. This ensures consistency with previous studies that have explored the influence of exotic degrees of freedom on the structure and stability of NSs. The inclusion of these additional particles plays a significant role in modifying the high-density EOS and, consequently, affects the global properties of NSs, such as their maximum mass and tidal deformability.

\begin{table} [t!]
\caption{The nuclear properties of the CDF models at respective $n_0$.}
\centering
\begin{tabular}{ccccccc}
\hline \hline
CDF Model & $n_0$ & $-E_0$ & $K_0$ & $E_{\text{sym}}$ & $L_{\text{sym}}$ & $m^*_N/m_N$ \\
 & (fm$^{-3}$) & (MeV) & (MeV) & (MeV) & (MeV) & \\
\hline
\vspace{0.1cm}
DD2 & 0.149065 & $16.02$ & 242.70 & 32.73 & 54.966 & 0.5625 \\
\vspace{0.1cm}
DD-ME2 & 0.152 & $16.14$ & 250.89 & 32.30 & 51.253 & 0.572 \\
\vspace{0.1cm}
DD-MEX & 0.152 & $16.14$ & 267.059 & 32.269 & 49.576 & 0.556 \\
\hline
\end{tabular}
\label{tab:2}
\end{table}

\subsection{Quasinormal oscillation modes}
Quasinormal modes (QNMs) represent the inherent oscillatory response of compact stars to disturbances affecting both their internal structure and the surrounding spacetime. The frequencies and damping times of these oscillations serve as key indicators of the NS’s composition and physical characteristics, offering valuable information about the nature of dense matter. These perturbations are typically expressed using spherical harmonic decomposition, which separates them into even and odd parity modes. In this study, we focus specifically on QNMs induced by fluid perturbations that couple to gravitational waves, thereby restricting our analysis to the dominant quadrupolar (\( l=2 \)), even-parity perturbations within the Regge-Wheeler metric framework \cite{1967ApJ...149..591T} given by,
\begin{equation}
\begin{aligned}
    ds^2&= -e^{2\Phi(r)}[1+r^lH_0(r)\mathcal{Y}_{lm}e^{i\omega t}]dt^2- 2i\omega r^{l+1}H_1(r)\mathcal{Y}_{lm}e^{i\omega t}dt dr \\ & + e^{2\Lambda(r)}[1-r^lH_0(r)\mathcal{Y}_{lm}e^{i\omega t}]dr^2 + r^2[1-r^lK(r)\mathcal{Y}_{lm}e^{i\omega t}][d\theta^2 \\ & + \sin^2\theta \ d\phi^2].
\end{aligned}    
\end{equation}

The spacetime perturbations in this approach are described by the metric functions \( H_0 \), \( H_1 \), and \( K \), which are associated with the perturbation amplitudes. These functions, in combination with the spherical harmonics \( \mathcal{Y}_{lm} \) and the complex QNM frequency \( \omega \), characterize the dynamical response of the star. The real part of \( \omega \) determines the angular frequency of the mode, while the inverse of its imaginary part defines the damping time, which quantifies how quickly the oscillations decay.  

The fluid perturbations inside the NS are governed by the Lagrangian displacement vector, represented in terms of the perturbation amplitudes \( W(r) \) and \( V(r) \) in the form,
\begin{equation}
    \xi^i= \{r^{l-1}e^{-\Lambda}W(r), -r^{l-2}V(r)\partial_\theta,-r^{l-2}\sin^{-2}\theta V(r)\partial_\phi\}\mathcal{Y}_{lm}(\theta,\phi) e^{i\omega t}. \label{eq:disp_vec_GR}
\end{equation}

The boundary conditions at the centre of the star are:
\begin{equation}
\begin{aligned}
    W(0) & = 1, \\
    H_1(0) & = \frac{lK(0)+8\pi(\varepsilon_0 + p_0)W(0)}{(l-1)(l+2)/2 + 1}, \\
    X(0) & = (\varepsilon_0 + p_0)e^{\Phi_0} \Bigl\{ \Bigl[\frac{4\pi}{3}(\varepsilon_0 + 3p_0) - \frac{\omega^2}{l}e^{-\Phi_0} \Bigr] W(0) +\frac{K(0)}{2} \Bigr\},
\end{aligned}
\end{equation}
where $\varepsilon_0$, $p_0$ are the local energy density and matter pressure at the centre of the star respectively. Another boundary condition at the surface of the star, $X(R)=0$, is obtained by considering two trial solutions, $K(0)=\pm (\varepsilon_0 + p_0)$.
At the stellar surface, fluid oscillations vanish, leaving only the perturbations of the surrounding spacetime, which propagate as gravitational waves. At asymptotic infinity, these perturbations can be decomposed into ingoing and outgoing gravitational wave components. The complex QNM frequency corresponds to the specific value of \( \omega \) for which only the outgoing gravitational wave component remains non-trivial, ensuring that the oscillation mode is purely radiative \cite{lindblom1983quadrupole,detweiler1985nonradial}.  

There exist multiple approaches for determining QNM frequencies, including resonance matching techniques \cite{1969ApJ...158....1T,chandrasekhar1991}, continued fraction methods \cite{sotaniDensityDiscontinuityNeutron2001}, and WKB approximations \cite{kokkotasWModes1992}. 
In cases where the perturbation of the gravity field or, the GR metric are not taken into account, leads to the relativistic Cowling approximation, i.e. considering $H_0=H_1=K=0$, one can obtain the relativistic Cowling equations \cite{1941MNRAS.101..367C}.
On the other hand, the direct numerical integration method \cite{lindblom1983quadrupole,detweiler1985nonradial,lujunli_ChinPhyB} requires solving linearized Einstein field equations alongside the fluid perturbation equations.
Even though the latter method is comparatively computationally expensive, it is very important for gravitational wave asteroseismological investigations. This is due to the fact that it provides the precise frequencies and damping times of oscillation modes needed to accurately interpret observational data.

In this work, we employ the method of direct numerical integration to obtain the oscillation frequencies and corresponding damping times. To classify different QNMs, we adopt the Cowling classification, which categorizes modes based on the number of radial nodes \cite{coxNonradialOscillationsStars1976,rodriguezThreeApproachesClassification2023}. 
The main basis of this classification scheme is on the displacement of fluid element from its equilibrium position inside the compact object when the restoring force (in the form of pressure or, buoyancy) acts on it \cite{1941MNRAS.101..367C}.
The modes are primarily defined based on the number of nodes in the radial eigenfunction.
Our primary focus is on the fundamental \( f \)-mode, which has zero radial nodes, and the first pressure \( p_1 \)-mode, characterized by a single radial node. 
It is noteworthy to mention that the pressure mode and the gravity ($g$)-mode has identical nodes. They can be distinguished based on their frequency comparison with the latter having a lower frequency.
To summarize, the pressure ($p_1-$) modes exhibit higher frequencies compared to the gravity ($g-$) modes, with the fundamental ($f-$) modes typically lying in between the two.
These modes are particularly significant as they provide valuable constraints on the NS's EOS and its overall compactness, thereby playing a crucial role in gravitational wave astrophysics.

\section{Results $\&$ Discussion} \label{sec:results}
In this section, we analyze the universal relations associated with compact astrophysical objects constructed using different dense matter EOS models. These relations establish EOS-independent correlations between various macroscopic properties of NSs, such as tidal deformability, moment of inertia, and quasinormal mode frequencies.
The EOS models considered in this study are chosen to be consistent with the NS observational constraints, ensuring their astrophysical viability. For a comprehensive discussion on the EOS models and their underlying assumptions, readers are encouraged to refer to Refs. \cite{PhysRevD.103.063004, 2021MNRAS.507.2991T}.  

The primary objective of this work is to investigate and validate the existence of universal relations in the context of QNMs. By exploring the correlations between oscillation frequencies, tidal deformability, and other macroscopic properties of NSs, we aim to establish whether these relations hold across different dense matter compositions, including exotic baryonic matter. This analysis provides further insight into the robustness of universal relations and their implications for NS astrophysics.

\subsection{$I-$Love$-Q$ relation}

We begin our analysis by investigating the $I-Love-Q$ relations, which involve computing the dimensionless parameters: the moment of inertia $\Bar{I} = I/M^3$, the spin-induced quadrupole moment $\Bar{Q} = -Q/M^3\chi_L^2$, and the tidal Love number $\Lambda = \Bar{\lambda} = \lambda/M^5$. Here, $I$ represents the moment of inertia, $Q$ denotes the spin-induced quadrupole moment \cite{1967ApJ...150.1005H, 2012PhRvL.108w1104P}, and $\chi_L = J/M^2$ is the dimensionless spin parameter, where $J$ corresponds to the total angular momentum of the star. These parameters are evaluated for NSs composed of pure nucleonic matter, hypernuclear matter, and $\Delta$-admixed hypernuclear matter. The calculations are performed using the RNS code \cite{1995ApJ...444..306S, refId0}, which numerically solves Einstein’s field equations for axially symmetric, stationary spacetimes in spherical coordinates. For consistency in the derivation of universal relations, the rotational frequency is fixed at approximately 480 Hz.
In order to calculate the moment of inertia and the quadrupole moment it is of importance to take into account the lowest possible rotation frequency, as the universal relations can be affected by rotational effects \cite{PhysRevD.99.043004}. Due to limitations of the RNS code, these quantities are computed at a fixed frequency of 480 Hz for all EOSs. Although the universal relations are known to deviate significantly near the Kepler frequency, the chosen frequency is much lower than the corresponding Kepler limit, ensuring minimal rotational influence.

\begin{figure}[t!]
    \centering
    \includegraphics[width=0.9\linewidth, keepaspectratio]{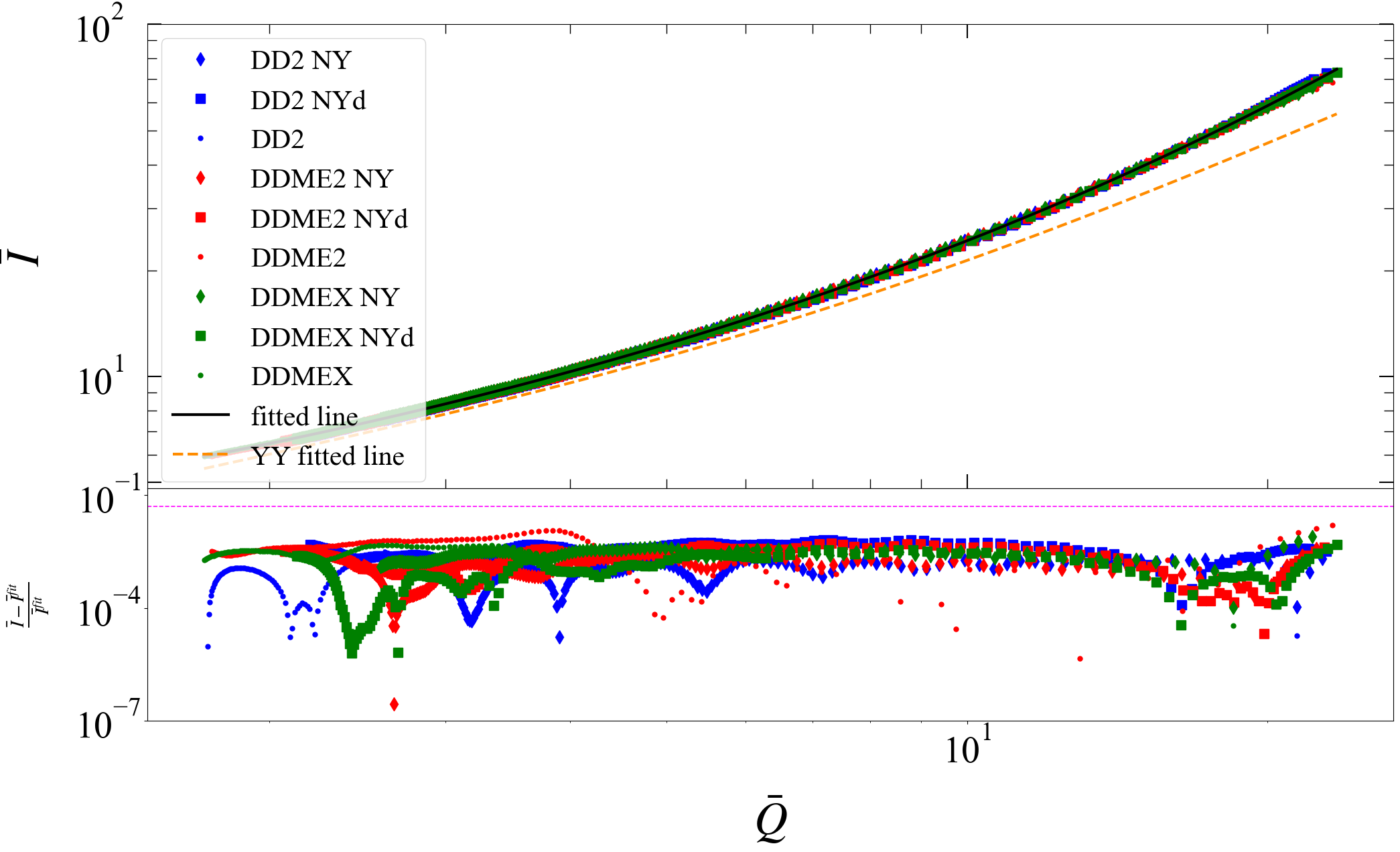}
    \caption{The $I-Q$ relations are presented alongside the corresponding analytic fit and the fractional error associated with the fitting function. The solid line represents the best-fit curve encompassing all data points. The inset plot illustrates the fractional errors, with the dashed line marking a 5$\%$ error threshold. The EOS compositions considered in this analysis are denoted as $NY$ for hypernuclear matter and $NYd$ for $\Delta$-admixed hypernuclear matter.}
    \label{fig:1}
\end{figure}

Figure \ref{fig:1} presents the correlation between the dimensionless moment of inertia ($\Bar{I}$) and the dimensionless spin-induced quadrupole moment ($\Bar{Q}$) for NSs modeled with different equations of state (EOS), as specified in the legend. The solid line represents the best-fit curve to the computed data points, while the dashed line corresponds to the empirical fit provided in Ref.- \cite{2013Sci...341..365Y}, commonly referred to as the YY fit.  

The universality in the $I-Q$ relation is captured using a fourth-order polynomial fit of the form: 
\[
\ln \Bar{I}_{\rm{fit}}=a_4(\ln \Bar{Q})^4+a_3(\ln \Bar{Q})^3+a_2(\ln \Bar{Q})^2+a_1(\ln \Bar{Q})^1+a_0
\]
where the coefficients of the fit are provided in Table \ref{tab:1}. The quality of the fit is assessed through the coefficient of determination, $\mathcal{R}^2 = 1 - SS_{\text{res}}/SS_{\text{total}}$, which evaluates how well the polynomial model describes the data. In this case, we obtain $\mathcal{R}^2 = 0.999$, indicating an almost perfect correlation between the fitted curve and the computed values. Here, $SS_{\text{res}}$ represents the sum of squared residuals, while $SS_{\text{total}}$ denotes the total variance in the data.  
To further assess the accuracy of the fit, the lower panel of Figure \ref{fig:1} displays the fractional error, comparing the computed values with the fitted curve. The maximum deviation is found to be within 5\%, reaffirming the robustness of the universal relation across different EOS models.  

An important observation is that the $I-Q$ relation obtained in our study exhibits noticeable differences at high quadrupole moments compared to the YY fit from Ref.- \cite{2013Sci...341..365Y}. This discrepancy arises primarily due to variations in the rotational frequencies assumed in the two studies. Since the moment of inertia and quadrupole moment are sensitive to rotation, the choice of rotational frequency plays a crucial role in determining their numerical values. In our analysis, we have adopted a specific rotation rate suitable for astrophysical applications, which differs from that considered in Ref.- \cite{2013Sci...341..365Y}, leading to slight variations in the extracted universal relation. Nonetheless, the overall trend remains consistent, demonstrating the robustness of the $I-Q$ relation as an EOS-independent feature of compact stars.

\begin{table*}[t!]
\centering
\caption{Coefficients of the tight correlations (polynomial fits) for different universal relations.}
\begin{tabular}{|c|c|c|c|c|c|c|c|}
\hline
Fits     & $a_5 (\times 10^{-7})$                  & $a_4 (\times 10^{-5})$                  & $a_3 (\times 10^{-3})$                  & $a_2$                  & $a_1$                  & $a_0$                     \\ \hline
$\ln{\Bar{I}}$ $-$ $\ln{\Bar{Q}}$ & 0 & 230 & 4.7  & 0.109  & 0.424  & 1.517 \\ \hline
$\ln{\Bar{I}}$ $-$  $\ln{\Lambda}$  & 0  & $-20$ & 4.1 & $-0.019$ & 0.21    & 1.342 \\ \hline
$\ln{\Bar{Q}}$ $-$ $\ln{\Lambda}$ & 0 & $-7.0704$ & $1.1$ & $-0.006$ & 0.324 & $-0.192$ \\ \hline
${\Bar{\omega}}_{f}$ $-$ $\ln{\Lambda}$ & $8.9866$ & $-4.2787$ & 0.8 & $-0.005$ & $-0.005$ & 0.181 \\ \hline
${\Bar{\omega}}_{p}$  $-$ $\ln{\Lambda}$ & 0 & 0 & $ 0.4$ & $-0.009$ & 0.01 & 0.437 \\ \hline
\end{tabular}
\label{tab:1}
\end{table*}

\begin{figure}[h!]
    \begin{subfigure}{0.9\textwidth}
        \centering
        \includegraphics[width=\linewidth]{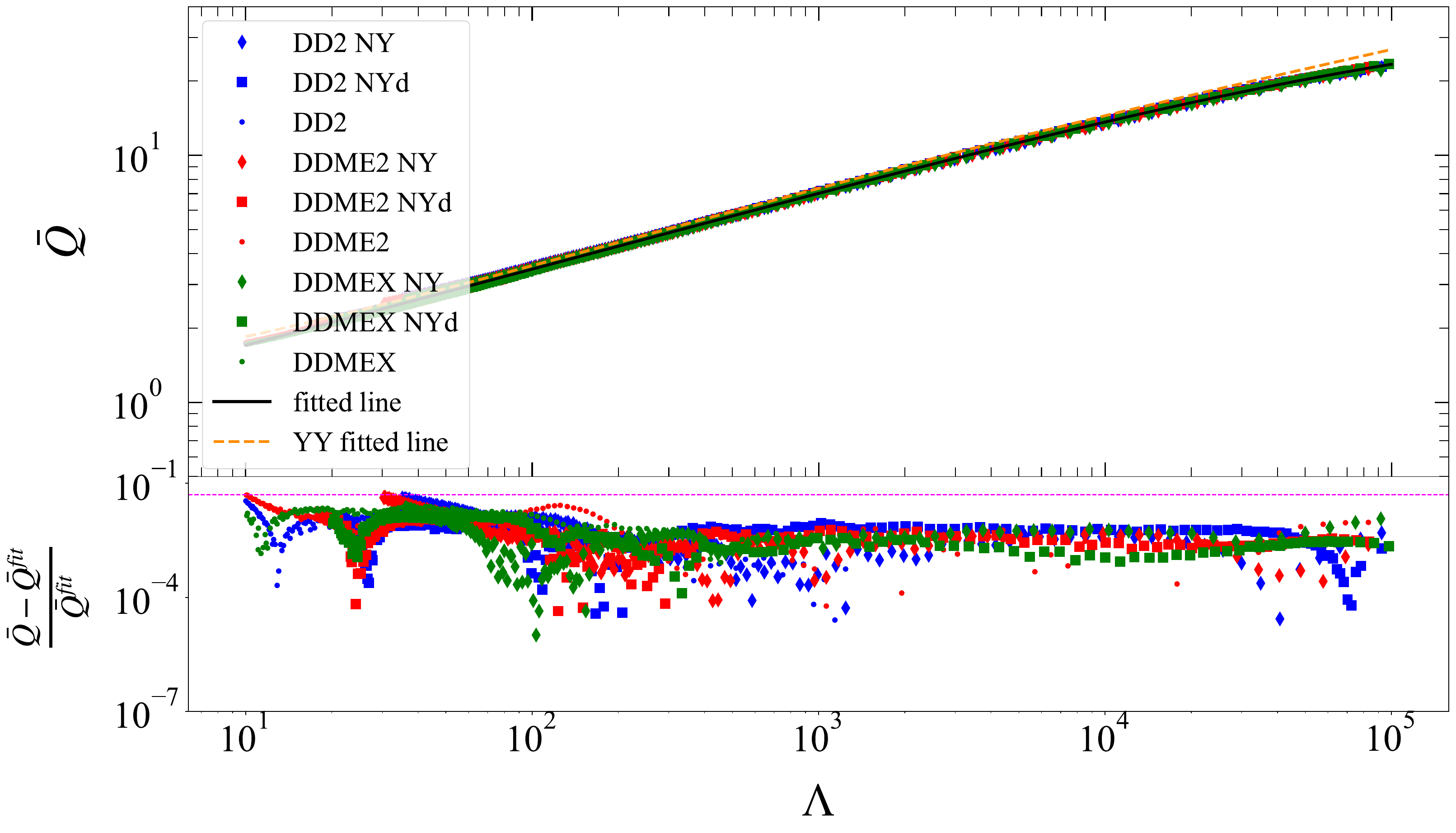}
    \end{subfigure}
    \begin{subfigure}{0.9\textwidth}
        \centering
        \includegraphics[width=\linewidth]{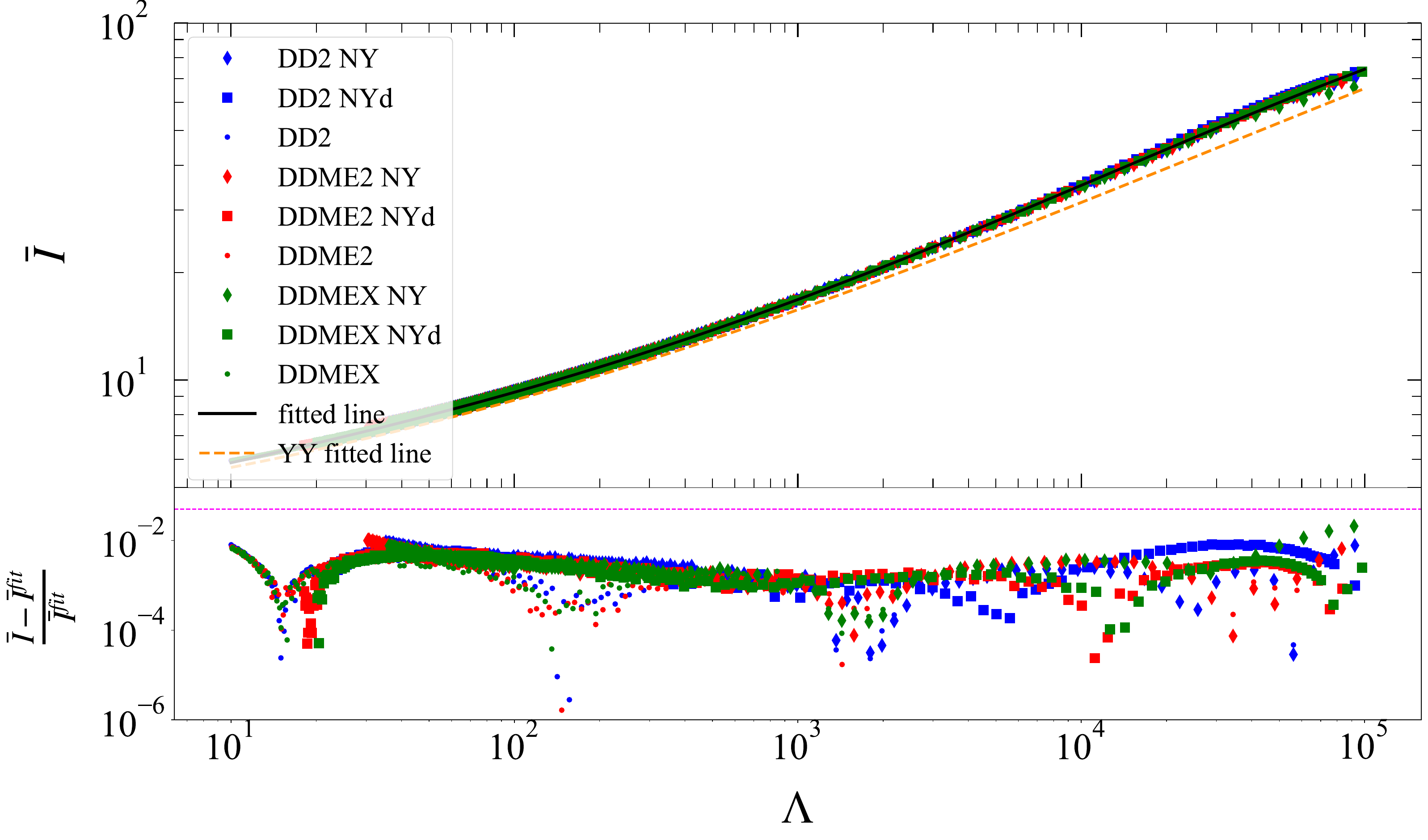}
    \end{subfigure}
    \caption{Universal relations for compact stars, with the upper panel depicting the correlation between the dimensionless spin-induced quadrupole moment ($\Bar{Q}$) and the tidal deformability parameter ($\Lambda$), while the lower panel presents the $I-\Lambda$ relation. The various EOS considered are indicated in the legend. The solid lines represent the polynomial fits to the data, with the corresponding fit coefficients provided in Table \ref{tab:1}. The sub-figures display the fractional errors, with deviations remaining within a few percent.}
    \label{fig:2}
\end{figure}

Figure \ref{fig:2} illustrates the universal relations between the dimensionless quadrupole moment ($\Bar{Q}$) and tidal deformability ($\Lambda$) in the upper panel, and between the dimensionless moment of inertia ($\Bar{I}$) and tidal deformability ($\Lambda$) in the lower panel. The solid lines represent polynomial fits to the computed data points, highlighting the strong correlations present in both cases. The coefficients of determination ($\mathcal{R}$) for these fits are exceptionally high, at 0.999 for both $\ln\Bar{Q}-\ln\Lambda$ and $\ln\Bar{I}-\ln\Lambda$, indicating minimal deviations from the fitted relations. The polynomial coefficients used for these fits are detailed in Table \ref{tab:1}. 

A quantitative analysis of the deviations from the fitted relations reveals a maximum discrepancy of approximately 4.65\% in the $\Bar{Q}-\Lambda$ relation, while for the $\Bar{I}-\Lambda$ relation, the deviation is even smaller, around 1.56\%. These small deviations confirm the robustness of the universal relations across different equations of state (EOS). Furthermore, the obtained trends align closely with those reported by Yagi and Yunes \cite{2013Sci...341..365Y}, reinforcing the idea that the $I-Love-Q$ relations hold universally, regardless of the underlying microphysics of NS interiors. This further validates the use of these relations as effective tools for probing the internal structure of compact stars using astrophysical observations.

\begin{figure*}[h!]
    \begin{subfigure}{0.9\textwidth}
        \centering
        \includegraphics[width=\linewidth]{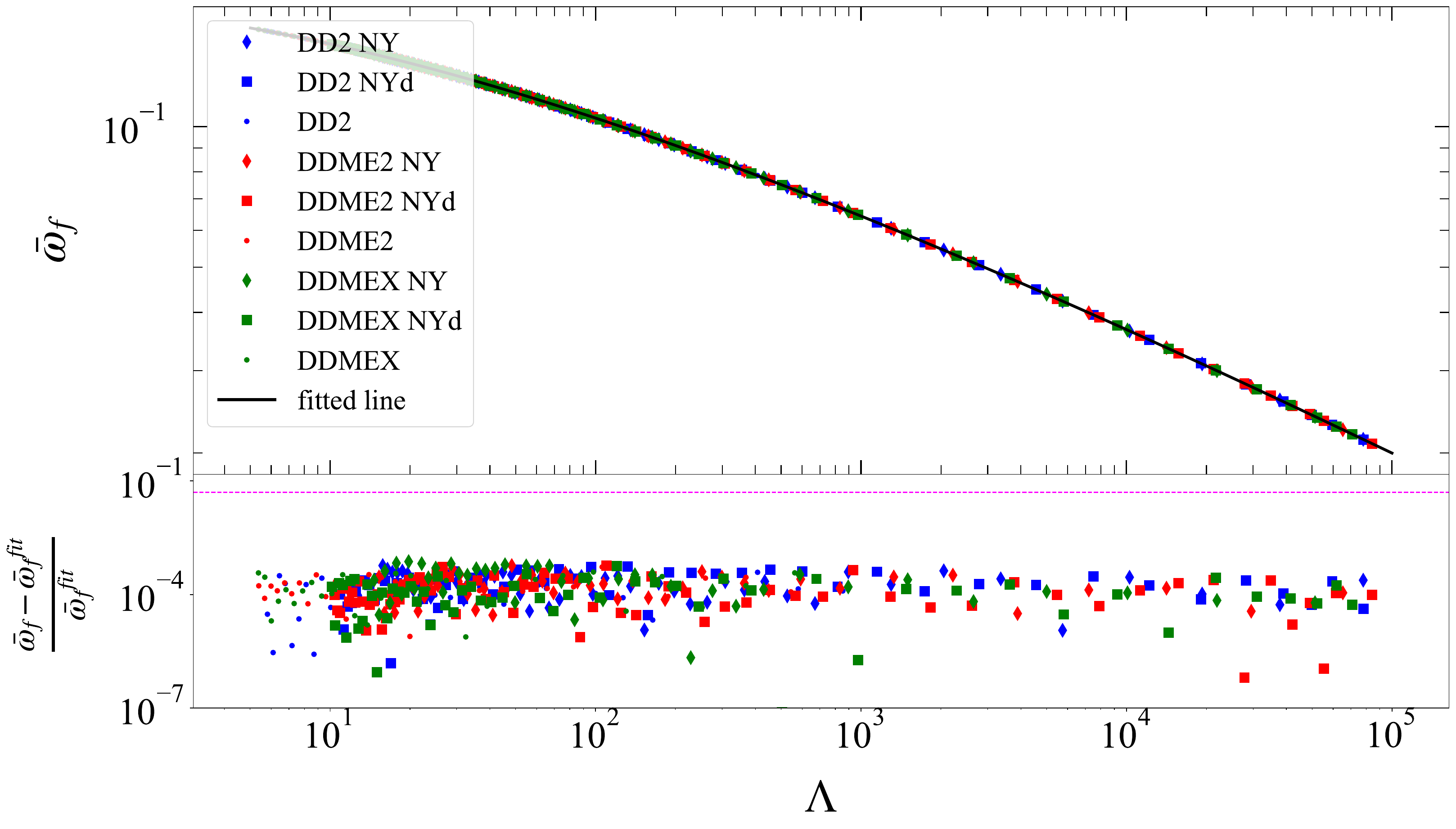}
    \end{subfigure}
    \begin{subfigure}{0.9\textwidth}
        \centering
        \includegraphics[width=\linewidth]{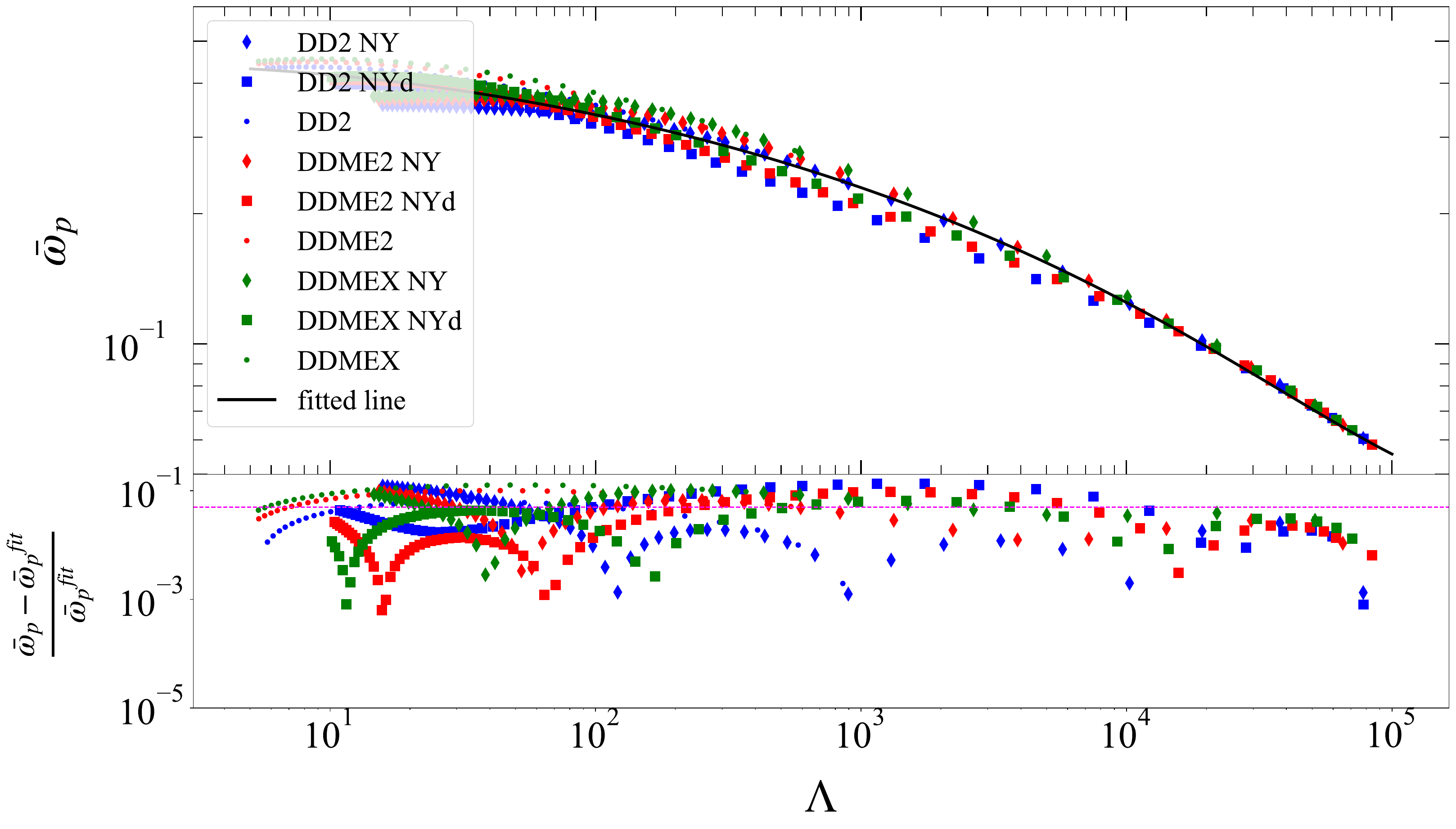}
    \end{subfigure}
    \caption{Similar to Figs.- \ref{fig:1} $\&$ \ref{fig:2}, but for the correlation fits for $\Bar{\omega}_f-\Lambda$ relation (upper panel) and for $\Bar{\omega}_p-\Lambda$ relation (lower panel).}
    \label{fig:3}
\end{figure*}

We now examine the correlations between the dimensionless frequencies of the fundamental ($\Bar{\omega}_f$) and pressure ($\Bar{\omega}_p$) quasi-normal modes (QNMs) with the dimensionless tidal deformability parameter ($\Lambda$). The QNM frequencies are computed using a fully general relativistic (GR) framework to ensure accuracy. Figure \ref{fig:3} presents the variation of the f- and p-mode QNMs with the tidal response for NSs composed of dense matter. 

For the fundamental mode ($\Bar{\omega}_f$), we obtain an exceptionally high coefficient of determination ($\mathcal{R} = 0.9999$) when using the full GR approach, confirming a strong correlation. In contrast, a slightly lower value ($\mathcal{R} = 0.9995$) is found when employing the Cowling approximation, although those results are not reported in this study. The maximum deviation from the fitted relation in the full GR case is only about 0.07$\%$, demonstrating the robustness of the universality in this relation. 
Additionally, previous studies, such as Ref. \cite{zhaoUniversalRelationsNeutron2022}, have established the universal nature of the $\Bar{\omega}_f-\Lambda$ relation for NSs composed of purely nucleonic matter. Our results extend this understanding by incorporating different EOS, further strengthening the argument that QNM frequencies can serve as effective probes of NS structure.

The right panel of Fig. \ref{fig:3} illustrates the variation of the pressure (p-mode) quasi-normal mode frequency with the dimensionless tidal deformability parameter ($\Lambda$). Unlike the fundamental (f-mode) case, the data points for the $\Bar{\omega}_p-\Lambda$ relation exhibit a wider spread, indicating a weaker correlation. 
The coefficient of determination for this fit is found to be $\mathcal{R} \approx 0.9684$, with a maximum deviation exceeding 14$\%$. A similar trend is observed when employing the Cowling approximation, reinforcing the conclusion that p-modes are more sensitive to variations in the EOS and matter composition compared to f-mode oscillations. 
This sensitivity of p-modes to dense matter properties has also been highlighted in Ref. \cite{thapaFrequenciesOscillationModes2023}, emphasizing their potential as probes of NS interior composition. The polynomial fit coefficients for both $\Bar{\omega}_f-\Lambda$ and $\Bar{\omega}_p-\Lambda$ relations are provided in Table \ref{tab:1}.

\section{Conclusions} \label{sec:summary}
At extreme densities, the properties of dense matter remain uncertain, both experimentally and theoretically. As a result, various high-density matter models have been proposed, incorporating different compositions and interactions among particles. The macroscopic properties of NSs, such as their structural parameters, are strongly influenced by the choice of EOS. However, despite these variations, certain parameter relations exhibit universal behavior, a concept first identified in Ref. \cite{2013Sci...341..365Y, 2013PhRvD..88b3009Y}. This universality has been extensively validated through subsequent studies \cite{2016CQGra..33mLT01Y, 2018CQGra..35a5005S, 2017PhRvC..96d5806M, 2013ApJ...777...68B, 2014PhRvL.112l1101P, 2019PhRvD..99d3004R, 2020PhRvD.101l4006J, 2021ApJ...906...98N, 2022MNRAS.515.3539K}, particularly for NSs composed of purely nucleonic matter.  

In this study, we extend the investigation of universal relations to dense matter configurations involving $\Delta$-admixed hypernuclear matter. Our analysis confirms that these exotic compositions adhere to the same universal relations as those established for purely nucleonic NSs. Specifically, the I-Q, I-$\Lambda$, and Q-$\Lambda$ relations exhibit strong correlations, with polynomial fits demonstrating excellent agreement with our numerical results. 

Furthermore, we explore the universality in the non-radial oscillation frequencies of NSs, particularly the $f-$ and $p_1-$mode QNMs. We find that the $f-$mode frequency ($\Bar{\omega}_f$) maintains a strong correlation with the dimensionless tidal deformability parameter ($\Lambda$), reinforcing its universality even in the presence of exotic baryonic matter. The maximum deviation in the fit remains minimal ($\sim 0.07\%$) in full general relativistic calculations. However, in contrast to f-modes, the p-mode frequency ($\Bar{\omega}_p$) exhibits a weaker correlation with $\Lambda$, with data points showing greater dispersion. The coefficient of determination for the $\Bar{\omega}_p$–$\Lambda$ relation is notably lower ($\mathcal{R} \approx 0.9684$), and deviations from the fit exceed 14$\%$. This suggests that p-modes are highly sensitive to the underlying matter composition, an observation consistent across both full general relativistic calculations and the Cowling approximation. 

Our findings reinforce the robustness of universal relations across different EOS models, including those incorporating exotic degrees of freedom. However, the observed sensitivity of p-modes to the composition highlights their potential as a diagnostic tool for probing the nature of dense matter in NSs.

\section*{Declaration of competing interest}
The authors declare that they have no known competing financial interests or personal relationships that could have appeared to influence the work reported in this paper.

\section*{Data availability}
The data set generated in this work can be provided on reasonable request to the corresponding author.



  \bibliographystyle{elsarticle-num} 
  \bibliography{references}






\end{document}